# A Privacy-preserving Central Bank Ledger for Central Bank Digital Currency

CHAN Wang Mong Tikvah

La Salle College

chan.tikvah@gmail.com

15 August, 2023

## Abstract

Central banks around the world are actively exploring the issuance of retail central bank digital currency (rCBDC), which is widely seen as a key upgrade of the monetary system in the 21$^{st}$ century. However, privacy concerns are the main impediment to rCBDC's development and roll-out. A central bank as the issuer of rCBDC would typically need to keep a digital ledger to record all the balances and transactions of citizens. These data, when combined with other data, could possibly disclose the spending habits of all citizens. On the one hand, the eligible rights of people to keep their transactions private should be protected, including against central bank surveillance. On the other hand, the central bank needs to ensure that no over-issuance of money or other frauds occur, necessarily demanding a certain form of knowledge of rCBDC transactions to safeguard against malicious users who create counterfeit money or spend duplicated money.

This work investigates cryptographic tools and privacy-enhancing technology with the aim to craft a scalable solution to strike a balance between user privacy and transaction verifiability. Different from the current mainstream thought among central banks, it assumes that the central bank maintains a ledger to record all balances and transactions of citizens, but in a concealed form. Specifically, this work focuses on rCBDC architectures based on the unspent transaction output (UTXO) data model and tackles the research problem of preserving a sufficient degree of privacy for UTXO transaction records while allowing the central bank to verify their correctness.

While UTXO-based rCBDC architectures were widely tested among major central banks, user privacy is not adequately addressed. The adoption of evolving public keys as pseudonyms to hide the real identities of users is the most advanced privacy design for UTXO-based rCBDC, but it only solves the privacy issue partially. Some information could still be leaked out. This work investigates techniques to address the shortcomings of the pseudonym approach.

First, a Pedersen commitment scheme is applied to hide the transaction values of a UTXO transaction while allowing the central bank to verify that no over-issuance of rCBDC has occurred in the transaction. Contrary to the conventional approach, which applies a zero knowledge proof to prove no over-issuance, this work uses a Schnorr signature. This not only reduces the overheads but also enables a non-interactive proof. Then, Coinjoin is applied to aggregate UTXO transactions from different users into one larger UTXO transaction to obfuscate the payer-payee relationship while preserving the correctness of the amount of money flow. This work applies a well-developed notion in database research, namely, k-anonymity, to analyse the privacy guarantee of Coinjoin. Through modelling the transaction traffic by a Poisson process, the trade-off between anonymity and transaction confirmation time of Coinjoin is analysed.

# Table of Contents

## 1)     Introduction

Central banks around the world are actively exploring the issuance of **central bank digital currency (CBDC)**.  The Atlantic Council's **CBDC** tracker reports that 130 countries, representing 98% of global GDP, are exploring **CBDC**[1].  Among these initiatives, **retail CBDC (rCBDC)** – a digital form of central bank money that is made available to the general public for payments – is widely seen as a key upgrade of the monetary system in the 21st century since physical cash (not digital) is the only, existing form of central bank money accessible to the public.  According to a recent survey by the **Bank for International Settlements (BIS)**, the responses from 86 central banks show that 93% of the respondents have been engaging in the development of **CBDC** and that the work on retail CBDC (**rCBDC**) is more advanced than on **wholesale CBDC (wCBDC)** [1].  Separately, the **International Monetary Fund (IMF)** claims that around 100 countries are exploring **rCBDC** [2].

Despite people's excitement, privacy concerns are the main impediment to the **rCBDC** roll-out [3-6].  User privacy is often seen as the most valued property and a key success factor that determines whether **rCBDC** would be generally accepted and used by the general public [4].  Many individuals feel that their financial transactions should be private, free from surveillance by any entity, including the central bank.  They consider that such privacy can support personal freedoms and protect sensitive information.  In fact, the low adoption rate of deployed **rCBDC** initiatives (including the Bahamas' Sand Dollar and Nigeria's eNaira) are often attributed to the insufficient privacy protection of the direct, one-tier **CBDC** model adopted by these initiatives, wherein, the central bank directly processes retail transactions and keeps the ledger of all transactions[2].

On the other hand, central banks have the mandate to maintain financial stability.  In other words, central banks need to control the money supply to manage inflation and ensure the overall health of the economy.  As the issuer of **rCBDC**, a central bank would therefore need to keep a digital ledger to record all the account balances and transactions of citizens in order to prevent over-issuance of money.  However, this repository for economy-wide transaction-level data, when combined with other data, could possibly disclose the spending habits of all citizens.  Besides, central banks consider the responsibilities for protecting privacy and user data, including from other arms of the government, onerous [8, 9].

---

[1] https://www.atlanticcouncil.org/cbdctracker/
[2] rCBDC can be provisioned through three different models, namely, the direct, hybrid and intermediated models.

In response, the intermediated model was proposed and adopted by many central banks [7, 8]. In the intermediated model, retail transactions are processed by commercial banks (or simply banks), and the central bank does not record any retail balance or transaction of citizens; the central bank only keeps wholesale balances, that is, the balances owned by banks on behalf of their clients. Without a ledger of citizens' account balances and transactions kept at the central bank, the citizens' privacy is thereby protected. However, the downside of this approach is that additional safeguards or oversight would be necessary, as the banks would need to be closely supervised to ensure at all times that the wholesale balances they communicate to the central bank accurately reflect the retail holdings of their clients. In addition, as [5] rightly points out, if a bank becomes insolvent, the central bank will have no data available to honor claims from citizens.

A real need therefore exists for central banks to record retail account balances and transactions of citizens. The intermediated model only skips tackling the privacy issue directly; it does not resolve or address the issue. The key to solving the problem is to strike the right balance between user privacy protection and sufficient disclosure for central banks to achieve their mandate. On the one hand, the eligible rights of people to keep their transactions private should be protected. On the other hand, the central bank needs to ensure that no over-issuance of money or other frauds occur, necessarily demanding a certain degree of disclosure of **rCBDC** transactions to safeguard against non-compliant banks which may incorrectly record account balances to create extra money supply, as well as, malicious users who create counterfeit money or spend duplicated money.

In view of this, this work tackles the privacy problem with a different approach, assuming that a ledger of retail balances and transactions (in a concealed form) is maintained at the central bank. In normal operations, these retail balances and transactions are inaccessible to the central bank but would give the central bank assurance that no over-issuance happens; whenever necessary and with the help from the concerned citizen (i.e., the respective data owner), the central bank can open the concealed balances and transactions. This approach is applicable to all **CBDC** models, including the direct, hybrid, and intermediated models, therefore widely applicable to different models adopted by central banks.

This work investigates different cryptographic tools with the aim to craft a scalable solution to strike a balance between user privacy and transaction verifiability. Simply encrypting the balances or transactions of citizens at the central bank ledger does not solve the problem. While it can protect citizen privacy, nobody can verify whether the payer of a transaction really pays the right amount in his transaction since all the information of the transaction is concealed and unverifiable by anyone (including the payee).

Specifically, this work focuses on **rCBDC** architectures based on the **unspent transaction output (UTXO)** data model [5, 10] as **UTXO** can support tracing of money supply and transaction flow to avoid over-issuance and counterfeit issues. User privacy is not addressed in [10]. A design with evolving pseudonyms is proposed in [5] to hide the real identities of users in **UTXO** transactions, partially solving the privacy issue. Nevertheless, some information could still be leaked. First, other information of a **UTXO** transaction, such as its transaction value, remains unconcealed. Second, the input and output relationship of a **UTXO** transaction and the inter-transaction linkage could reveal or infer important information about the users, say, through analysing the transaction graph.

In detail, the solution of this work builds on the evolving pseudonym approach of [5] and proposes to complement it with two new mechanisms to conceal transaction information and obfuscate the payer-payee relationship respectively to address the shortcomings of the evolving pseudonym approach.

1) In the first mechanism, transaction values of the inputs and outputs of a **UTXO** transaction are concealed by a **Pedersen commitment scheme** [11]. The commitment scheme hides the transaction values of the **UTXO** inputs and outputs while supporting additions and subtractions to be carried out over them in the concealed form. When combined with a suitable ***zero knowledge proof*** [12], it can prove to the central bank that the **UTXO** transaction is properly formed without any over-issuance of **rCBDC** introduced in the process. But instead of combining the commitment scheme with a conventional zero knowledge proof, this work proposes to use a **Schnorr signature** [13] to construct a more efficient and non-interactive protocol for verifying no over-issuance in a concealed UTXO transaction.

2) In the second mechanism, called the **Coinjoin** [14, 22], multiple **UTXO** transactions from different citizens are aggregated or merged into a larger **UTXO** transaction to obfuscate the input-output relationship of the resulting **UTXO** transaction. In this way, when combined with the transaction value concealment of the first mechanism, it becomes less easy to ascertain who pays whom when seeing just the resulting **UTXO** transaction since a given payee could have been paid by any one of the payers in the merged transaction. Similarly, it is not likely to tell which one of the listed payees a given payer has paid to. Unlike the Coinjoin design of **MimbleWimble** [14, 22], this work does not take away the payer/payee public keys from the inputs/outputs of a **UTXO** transaction. While the number of transactions that can be merged together in this type of transaction aggregation is smaller compared to MimbleWimble, a proper payment authorization by respective **rCBDC** owners is preserved instead. Besides, the level of transacting counterparty anonymity of the proposed design is tunable. This work also links Coinjoin with a well-known notion in database research, namely, ***k-anonymity*** [15], thus allowing the anonymity of transaction aggregation to be properly analysed in a well-defined framework. Through modelling the transaction traffic with a Poisson process, the trade-off between anonymity and transaction confirmation time needed to wait for other transactions to form an aggregated transaction is studied.

The contribution of this work is two-fold. First, it introduces a design to enhance the transaction privacy of **UTXO**-based **rCBDC**. While concealing information of a transaction, the design also allows the central bank to verify that no over-issuance of **rCBDC** has occurred in the concealed transaction. Cryptographic protocols used are analysed. It also improves counterparty anonymity through the aggregation of multiple transactions into a larger one. The resulting protocol can readily support all **CBDC** models proposed by central banks. Second, this work lays out an evaluation framework for **CBDC** counterparty anonymity and demonstrates the possible design trade-off, thereby allowing performance tuning to be done in practice. Both analytical and simulation results will be presented.

## 2)   Literature Review

Various central banks proposed using **UTXO** in a centralized database as the ledger for **rCBDC** [3, 5, 10]. The privacy mechanism of this work is based on the same model. In the **UTXO** data model, instead of arranging the ledger as a list of pairs of accounts and balances, the ledger consists of a list of **UTXO**s (which are units of variable amounts of money) and their respective owners. That is, a user could own multiple **UTXO**s on the ledger. When a user spends his money, he specifies the **UTXO**s to be spent such that these **UTXO**s will be erased from the updated ledger and replaced by new **UTXO**s spending them. As [5] points out, the advantage of the **UTXO**-based model for **CBDC** is its traceability of money flow, ease of audit, and efficiency for implementing pseudonyms.

According to [7, 8], the intermediated **CBDC** model was proposed to avoid keeping a ledger of retail balances and transactions of citizens at the central bank, thereby precluding the possibility that the central bank or other government departments can access citizens' transaction data and invade their privacy. However, this approach poses various issues including a greater supervisory burden on the part of the central bank, a lack of data to verify that no over-issuance of money occurs, and no data at the central bank to honor claims from citizens if a bank becomes insolvent.

Various pseudonym designs were proposed in **CBDC** [5, 16]. The basic idea is to use randomly selected numbers, which can be public keys, instead of the real identities of users, in transactions recorded on the central bank ledger. [5] proposed using a new public key for each transaction to reduce the identifiable links between multiple transactions of the same user. This partially achieves the effect of mixing [18]. To ease key management, [5] proposed deriving new private keys through key derivation using a pseudorandom function. However, other information of a transaction, such as the transaction value, is not concealed. In some cases, certain transactions belonging to the same owner could still be linked, and the payer-payee relationships are known. The mechanisms designed in this work aim to address these shortcomings.

Outside the **CBDC** context, various privacy mechanism designs have been proposed for cryptocurrencies, for both **UTXO**- and account-based blockchain networks. However, in general, there are several limitations to directly apply these designs to **CBDC**, and complexity exists to tailor these designs for use in **CBDC**. First, many of these designs aim for a very high level of privacy without any information leakage, which might create problems when deployed in **CBDC**. As explained earlier, a real need exists for central banks to have access to certain information about user transactions to fulfill their mandates, such as for verifying that neither the banks nor users have created extra money supply to cause inflation. These privacy designs fail to offer a desirable handle for central banks to fulfill their mandates or meet regulatory requirements. Second, these privacy designs, in one way or another, rely on the underlying blockchain to build their privacy protection capability. Complexities may arise when they are applied directly to a centralized database setting, which is a mainline approach for **rCBDC** [3, 5, 10].

**Zerocash** [18] and **Zerocoin** [19] are protocols that provide Bitcoin (also a **UTXO**-based design) with strong privacy guarantees. In both protocols, instead of using Bitcoins to transact directly, Bitcoins are deposited into escrow to mint a new type of coins, called Zerocoins, which are the actual instrument used for transactions. Each Zerocoin is used once, thereby leaving no transaction history for tracing. In Zerocash, Zerocoins have hidden values and owners, therefore offering anonymity and privacy protection. A **cryptographic hash function** and a **Merkle tree** are used to create a cryptographic commitment, which serves as proof of ownership of the minted Zerocoins in Zerocash. When a Zerocoin is spent or consumed, the respective locked Bitcoins in the escrow are released while hiding the origin, destination, and amount of the transaction. To ensure that no extra Bitcoins are spent, a zero knowledge proof, specifically, *zk-SNARK*, is used to prove that the transaction preserves the total value of coins and prevent double-spending. Zerocoin is an earlier version of Zerocash, which only hides the origin of a payment transaction and uses different cryptographic tools to implement the commitment and zero knowledge proof.

The **anonymity voucher protocol** [17] uses a protocol similar to Zerocoin and Zerocash to escrow **CBDC** to create 'anonymity vouchers', which allow users to anonymously transfer a limited amount of **CBDC** over a defined period of time. The protocol has to build on **Distributed Ledger Technology** (**DLT**). As a result, it might not be directly applicable to the centralized database setting under consideration by various central banks.

Various mixing protocols [20, 21] have been proposed for Bitcoin. A mixing service collects Bitcoins from multiple users and sends them back in random amounts to new addresses controlled by the users. In this way, no one seeing the transactions can tell who has exchanged Bitcoins with whom, or how much they have exchanged. In essence, this shuffles the Bitcoins among the participating users to cut off the history and the provenance of the Bitcoins received in the new addresses. While providing a good degree of anonymity, it might not be trivial to deploy a mixing service in the centralized database setting of **CBDC**. The Coinjoin mechanism discussed in this work also achieves a partial mixing effect, and this work proposes to use the $k$-anonymity notion and a Poisson process to analyse the anonymity of Coinjoin and demonstrate the trade-off involved.

Mimblewimble [22] is a protocol that hides the details of **UTXO** transactions using three techniques, namely, Confidential Transactions, Coinjoin, and block aggregation. Confidential Transactions hide the transaction amounts through a Pedersen commitment scheme using random values that only the payees know. Coinjoin mixes multiple transactions together to obscure their origin and destination. Block aggregation combines all the inputs and outputs of the transactions occurring in the same block into one large transaction with possible input-output cancellation, making it impossible to link or trace them. These techniques make Mimblewimble transactions more private and scalable than Bitcoin transactions. One major disadvantage of Mimblewimble is that the payee of a Confidential Transaction needs to be online as it is an interactive protocol executed between the payer and payee to hide the transaction amount. In contrast, in this work, the protocol to hide the transaction value does not require the payee of a transaction to be online and is basically non-interactive. In MimbleWimble, there is no limit on the number of transactions that can be aggregated via Coinjoin.

In contrast, this work requires the payers of different transactions to sign on the same Coinjoin transaction, therefore limiting the size of aggregation. This work and MimbleWimble have different focuses. While MimbleWimble aims to maximize privacy protection, completely eliminating the use of public keys for both payers and payees, this work retains user public keys in transactions and strives to clearly delineate accountability and allow sufficient transaction evidence that can be presented to court whenever needed, which is an essential basis for CBDC.

Other works exist that use a ring signature scheme to protect the anonymity of users [23]. However, due to the challenges posed for regulatory compliance, these schemes are unsuitable for use in **CBDC**.

### 3)    Problem Setting

This work is largely based on the **rCBDC** architecture of [5], building on the design of **UTXO**-based transactions in the centralized database setting and using evolving user public keys as owner identities of **UTXO**s. The setting of **UTXO** transactions in a centralized database is chosen in this work because it is the model widely explored by major central banks including the Federal Reserve and European Central Bank. In the architectural model of [5], commercial banks (or simply banks) serve as intermediaries between the central bank and citizens to process retail transactions. The central bank does not process or record retail transactions so as to protect user privacy (against the central bank and other government departments). [5] argues that, based on a survey conducted by the **BIS** [24], citizens tend to trust banks to safeguard the privacy of their transaction data. In contrast, this work (while still assuming the banks acting as intermediaries to process retail transactions) assumes that the central bank ledger is the ultimate source of reference for all parties (including the central bank, banks and citizens) which records all transactions, including retail transactions initiated by citizens. That is, in this work, the central bank records retail transactions and balances of citizens (but in a concealed form).

The rationale of keeping all transaction data on the central bank ledger is two-fold. First, the central bank would be a trustworthy party to maintain the record of all transactions to prevent double-spending and other frauds as this is in line with its mandate to maintain financial stability and avoid over-issuance of money. This would also give the central bank data to verify that no over-issuance of money or double spending occurs in the retail layer. Second, to protect citizen privacy, this work believes that technology could help solve the problem. That is, technology could be used to protect user privacy, instead of depriving central banks of the data they need to fulfill their mandates. In fact, this work proposes the use of two privacy enhancing mechanisms to safeguard user privacy while allowing the central bank to keep all retail transactions.

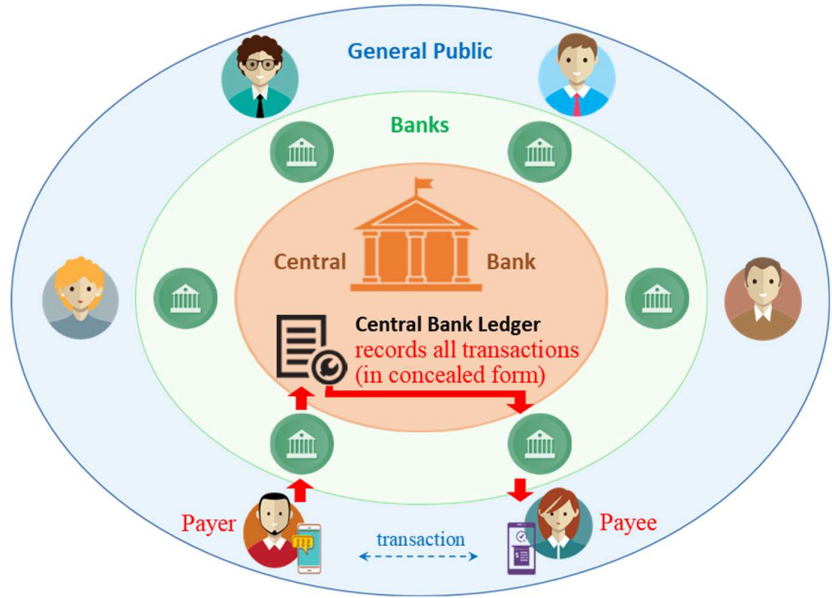

Figure 1.  **rCBDC** architecture and transaction flow with banks as intermediaries

Following [5], this work assumes that the mapping between user public keys (which are used as pseudonyms in **UTXO** transactions) and real user identities should be kept by banks.  That is, while a **UTXO** record at the central bank can tell that the ownership of the **UTXO** or **CBDC** belongs to a certain public key, it will not reveal to the central bank the identity of the actual person who owns the **UTXO** or **CBDC**.

### 3.1)    UTXO transactions

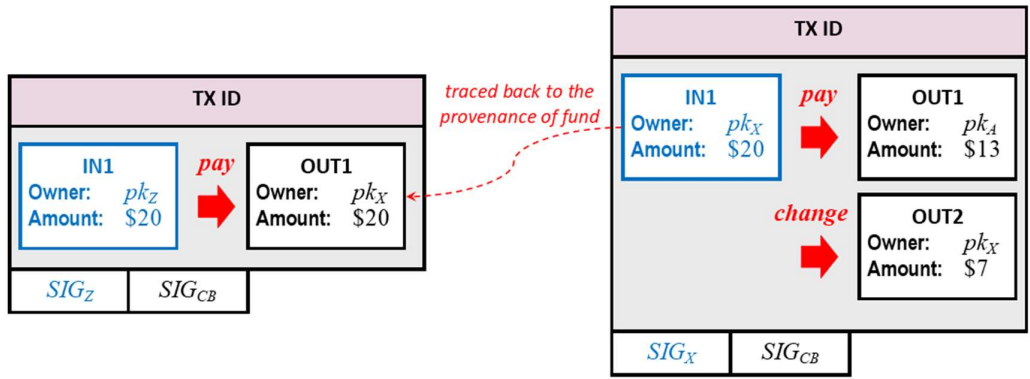

Figure 2. A typical **UTXO** transaction

The privacy design of this work is based on the **UTXO** transaction format. Figure 2 depicts the format of a typical **UTXO** transaction. In the **UTXO** data model, instead of recording citizens' ownership of **CBDC** as a list of accounts and the respective balances, the ledger records **CBDC** ownership as a list of different units of **CBDC** of variable amounts and the respective owners. Each of these units of **CBDC** is called a **UTXO**, with its amount and owner recorded as an entry on the ledger. Only the marked owner of a **UTXO** is authorized to spend it. Note that a citizen could possibly own multiple **UTXO**s (i.e., multiple entries) recorded on the ledger whereas each citizen would normally own one account on the ledger.

To create a transaction for payment, a payer needs to collect the information of a number of **UTXO**s recorded on the ledger whose total amount or value suffices to cover the payment. For example, to make a payment of $20, a payer could choose from the ledger records a **UTXO** with an amount of $20 or two **UTXO**s each corresponding to $10. In the example shown in Figure 2, a single **UTXO** of $20 is used by a citizen $X$ owning public key $pk_X$ to make payment to another citizen $A$ owning public key $pk_A$ (i.e., the payee). The selected **UTXO** to be spent is put as an input of a payment transaction. The outputs of the transaction correspond to the units of **CBDC** to be paid to the payees. Each of these outputs is a **UTXO** to be recorded on the ledger (after the transaction is confirmed), therefore containing information of the new owner of the respective unit of **CBDC** and the amount of **CBDC** owned.

The values of **UTXO**s recorded on the ledger may not always closely match the amount in payment. In such cases, a change could be made. As shown in Figure 2, the payer $X$ only needs to pay the payee $A$ an amount of $13, which is recorded in the first output **OUT1**, along with the public key $pk_A$ belonging to the new owner of the **CBDC** unit (i.e., $A$). The change of $7 is recorded in the second output **OUT2**, along with the payer's public key $pk_X$, as the **UTXO** is paid back to the payer as change. It should be noted that, for a valid transaction, the sum of all outputs should be equal to the sum of all inputs.

In order to authorize the payment to the payee $A$, the payer $X$ signs on the transaction, and $SIG_X$ is the signature of $X$ whose validity can be publicly verified by anyone using the payer's public key $pk_X$. The central bank verifies the validity of the signature $SIG_X$ before the change of ownership of the **rCBDC** is confirmed and recorded on the central bank ledger.

Upon the receipt of this transaction, the ledger operator (i.e., the central bank in this case) checks whether all the inputs of the transaction can be found in the latest record of the central bank ledger. If all can be located on the ledger, the ledger operator will erase all the **UTXO**s corresponding to these inputs and add the outputs of the transaction as new **UTXO**s to the ledger. In this way, the ownership of the **CBDC** units on the ledger is updated. To confirm that the transaction has been processed with the respective ledger records updated, the ledger operator signs on the transaction. Through verification of this signature $SIG_{CB}$, the payee can confirm that he has received the payment because the ownership has been updated on the central bank ledger (the authoritative source of records for all parties to reference). Subsequently, a payee can spend the received money by locating amongst the newly added **UTXO**s the one owned by him and using it as input of a new transaction.

Since an input of any transaction points back to an output of an earlier transaction, with transaction history, one can trace back the flow of money by linking the transactions together in a chain. In this way, the central bank can verify that the amount of money flow is preserved in the chain of transactions since the issuance of the money (i.e., no over-issuance as the ownership of the money changes).

### 3.2) Pseudonymous transactions

Since no real identities are used in transactions, the transactions are pseudonymous. However, since **UTXO** transactions are traceable, it is possible to determine where a unit of **CBDC** originates through analysing the transaction chain. In fact, it was demonstrated that **UTXO** transaction graph analysis can disclose more information than people expect in the case of Bitcoin [25]. Therefore, [5] proposed to use evolving public keys to increase the difficulty to infer information about the links between different transactions. To meet regulatory requirements, [5] proposed to maintain the mapping between these public keys and the real identities of users at banks, which can be revealed when necessary and upon proper authorization by the court. This work will adopt the same approach to manage the mapping between evolving public keys and real identities.

Depicted in Figure 3 is the detailed setup of the **rCBDC** architecture and the transaction processing flow adopted in this work. While the discussion in this work assumes the bank as an intermediary to relay transactions and messages between the central bank and user, the proposed design also works without relying on the bank if the user's device takes up all the computation (i.e., the direct **CBDC** model). In other words, the proposed privacy design would be applicable to all **CBDC** models, including the direct, hybrid and intermediated models.  This greatly increases the applicability of the proposed privacy solution.

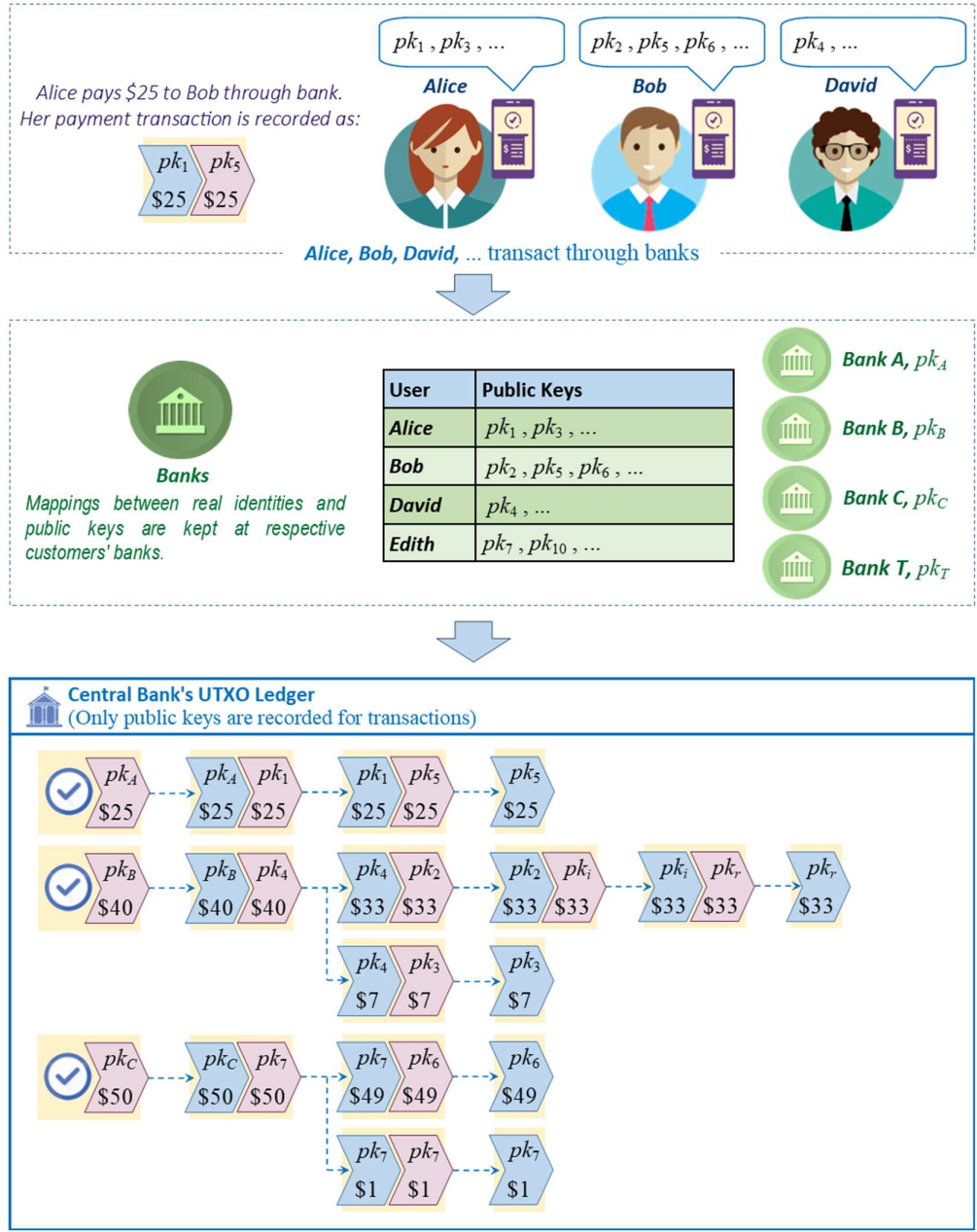

Figure 3.  **rCBDC** architecture and transaction processing flow

### 3.3)    Remaining privacy challenges

While [5] has considerably improved the privacy design of **rCBDC**, two challenges remain that might stir up privacy concern or fear [6] among citizens, possibly with impact to jeopardize the development and roll-out of **rCBDC**.

1) The transaction amount of value remains in clear text in the design of [5], meaning that the central bank would be able to see this information. In some case, say, a specific amount to be commonly used for a particular purpose, knowledge of the transaction value could help infer the nature of a transaction or payment.

2) Despite that the identities of the payer and/or payee of a transaction are not revealed to the central bank, the payer-payee relationship of the transaction is revealed to the central bank. Through the inter-transaction linkage of **UTXO** transactions, it might be possible for the central bank to infer additional information about the payer and payee. For instance, side information may be combined with the transaction graph to gain additional information. In other cases, the identity of the payee, say, a popular chain store, may be known to the public, the transaction information available may possibly infer the spending habit of a user. For instance, a user may be found frequently making payments for gambling purposes.

It is understandable why the design in [5] leaves this information of transactions unconcealed. As the central bank needs to assure that no over-issuance of money or inflation occurs in any given transaction, it would need to trace the flow of **rCBDC**, and knowing the transaction value and the source of fund of a transaction is therefore critical for the central bank to fulfill its mandate. However, this work is of the view that the disclosed information is more than necessary for the central bank to verify that no over-issuance of money occurs in any transaction. This work will show how cryptographic tools and privacy enhancing technologies could help the central bank to meet its objective while not invading the transaction privacy of citizens. This work is based on two intuitions as follows:

1) The central bank only needs to verify that the total amount of the outputs of a **UTXO** transaction is equal to the total amount of its inputs. In this way, it can be assured that no over-issuance has occurred in the transaction. Knowing the exact values of these inputs and outputs are unnecessary.

2) The central bank only needs to verify that the aggregate flow of **rCBDC** has no over-issuance or no new unit of **rCBDC** is created in the aggregate flow. It does not have to know, at a granular level, which transaction funds which transaction as long as it can verify that the overall fund inflow of a group of transactions is equal to the outflow. That is who pays whom is not necessary to verify that the total amount of money flow is preserved in a group of transactions.

Based on the first intuition, cryptographic tools [11, 13] are explored to conceal the transaction value of a **UTXO** transaction while allowing the central bank to verify that no over-issuance has taken place in the concealed transaction values. Based on the second intuition, a privacy enhancing technique called the Coinjoin [14, 21, 22] is explored to aggregate multiple transactions from different payers into a larger transaction. When combined with transaction value concealment, the payer-payee relationship could be obfuscated.

### 4)     Concealing Transaction Values

One of the objectives of this work is to conceal the transaction value of a **UTXO** transaction while allowing the central bank to have sufficient information to confirm that no over-issuance of **rCBDC** has occurred in the transaction. Recall that the central bank would only keep a ledger of concealed **UTXO** transactions from all the citizens. A bank has to prove to the central bank, on behalf of its client, that a given **UTXO** transaction in the concealed form is correctly formed without over-issuance of money, that is, the sum of the transaction values of all its outputs is equal to the sum of the transaction values of all its inputs.

A trivial approach to solve this problem is to let the bank encrypt the transaction values of all the inputs and outputs of a **UTXO** transaction, and then prove to the central bank in the encrypted form that the sum of these outputs is equal to the sum of these inputs, using a standard zero knowledge proof. However, we will show below that this approach may not adequately address the objective. First, running a zero knowledge proof over encrypted values is very inefficient in terms of computational and communication overheads. Even with $zk$-SNARK [12], the overhead could still be large. More importantly, encrypting the transaction values may not be secure as encryption does not necessarily guarantee the integrity of the plaintext value in the encryption. Namely, an encryption scheme does not necessarily guarantee that a given ciphertext will always be decrypted to recover exactly the same plaintext that has been used to prove to the central bank that no over-issuance has occurred. This is because, for a probabilistic encryption scheme, multiple plaintexts could be encrypted to give the same ciphertext. Below will show how a possible vulnerability may occur at redemption if a certain encryption scheme is used to conceal transaction values.

***Example 4.1*** Take **AES-counter mode (AES-CCM)** as an example.

For **AES-CCM**, the ciphertext $c = f(k; counter) \oplus m$, where $m$ is the message to be encrypted, $k$ is the secret key, $f$ is AES block cipher (treated as a pseudorandom function) and counter = 0, 1, 2 … is the counter/block value which is fixed and publicly known. Without loss of generality, assume the length of $m$ is 128 bits (i.e., the block size of AES). Then counter = 0.

Suppose the central bank is given an **AES-CCM** encrypted value for recording on its ledger:
$$X = E_k(q_1) = f(k; counter) \oplus q_1$$
under encryption key $k$ for the transaction value $q_1$. Note that $k$ is selected by the bank and unknown to the central bank. Suppose the bank receives a payment from its client and $q_1$ is the value it receives. Now the bank manages to prove to the central bank the validity of the payment transaction (i.e., the amount of monetary values is preserved in the transaction) through a zero knowledge proof. That is, $X$ is recorded on the central bank ledger as a valid **UTXO** owned by the bank. Later on, when the bank redeems the **CBDC** at the central bank, say, to top up its wholesale account, it needs to reveal $k$, $q_1$ to the central bank to open $X$ to prove to the central bank the amount to be redeemed. To verify, the central bank encrypts to see if it gets back $X$. If yes, it assumes $X$ is the encryption of $q_1$ and pays the bank accordingly.

If the bank manages to find another pair $(k', q_1')$ which encrypts to the same $X$, the central bank will accept $q_1'$ as the amount owned by the bank in $X$ instead. A malicious bank can try different pairs of $(k', q_1')$ until:
$$X = E_{k'}(q_1') = f(k'; counter) \oplus q_1'$$
If $q_1' > q_1$, the bank manages to redeem extra money than it actually owns.

Assuming $f$ performs like a random function. Then $f(k'; counter)$ is like a 128-bit random number as we pick different $k'$ randomly. Since $X$ is fixed, $q_1'$ would be like a number randomly picked from all 128-bit numbers. The probability that the bank can find a $q_1'$, such that $q_1' > q_1$, is given by $\dfrac{2^{128} - q_1 - 1}{2^{128}}$ and such probability decreases as $q_1$ increases.

It is rather dangerous as a larger proportion of transactions tend to have lower transaction amounts, leading to the probability that the bank redeeming extra money is relatively high.

As can be seen, encryption may not suffice to achieve the purpose of concealing the transaction values while allowing a certain degree of verifiability by the central bank. Instead, a Pedersen commitment scheme is used to conceal the transaction values of the inputs and outputs of a **UTXO** transaction. The reason why a Pedersen commitment scheme does not have the problem presented by an encryption scheme is that it is binding, namely, the value concealed in a commitment cannot be changed by the creator once the commitment is disclosed.

### 4.1)    Pedersen commitment scheme

A Pedersen commitment scheme [11] works over any multiplicative group, such as $Z_p{}^*$, a multiplicative sub-group of $Z_p{}^*$, or even points over an elliptic curve.  Note that $Z_p$ is integer mod $p$ where $p$ is a large prime number.  A Pedersen commitment scheme for a multiplicative subgroup of $Z_p{}^*$ of order $q$ involves a sender and a receiver and works as follows:

First, the receiver chooses large primes $p$ and $q$ such that $q$ divides $p-1$, a generator $g$ to construct the subgroup of $Z_p{}^*$, a random integer $a$ from $Z_q{}^*$. The discrete log problem of the chosen subgroup is supposed to be computationally hard.

The receiver then calculates $h = g^a \bmod p$.  Note that $a$ is hidden from the sender.

The values $p$, $q$, $g$, $h$ are public, while $a$ is kept private.

To hide a number $x$ taken from $Z_q{}^*$, the sender then chooses random $r$ from $Z_q{}^*$ and sends $c = g^x h^r \bmod p$ to the receiver.

To open the commitment, the sender sends $r$ and $x$ to the receiver, the receiver then verifies if

$$c = g^x h^r \bmod p .$$

A Pedersen commitment scheme has two interesting properties, namely, hiding and binding.

It is hiding in the sense that, given a commitment value $c = g^x h^r$, it is not possible for anyone to tell from $c$ what $x$, $r$ are.  In fact, Pedersen commitment achieves better assurance.  It is not possible to tell apart the actual commitment value from a truly random number.

It is binding in the sense that, once the sender uses $x$, $r$ to generate the commitment value $c$ and shares it with the receiver, the value $x$ is committed, and he cannot change its value without changing $c$. More precisely, once he commits $(x, r)$ to generate $c$, it is not possible for him to find another pair $(x', r')$ such that $(x', r')$ would give the same commitment value $c$.  If he manages to find the pair $(x', r')$ such that $g^{x'} h^{r'} = c \bmod p$, he can solve the discrete log problem in the chosen multiplicative subgroup of $Z_p{}^*$, which is computationally hard.  In detail,

$$g^{x'} h^{r'} = g^x h^r = c \bmod p$$
$$g^{x'+ar'} = g^{x+ar} \bmod p$$
$$(x' + ar') = (x + ar) \bmod q$$
$$a = \frac{x - x'}{r' - r} \bmod q$$

Note that he knows $(x, r)$ and $(x', r')$. However, solving for $a$ (without knowing it) is equivalent to solving the discrete logarithm of $h$, which is assumed to be hard to solve.

In other words, if the discrete log problem is computationally difficult for the given multiplicative group, then Pedersen commitment is binding.

Pedersen commitment has a very useful property, namely, the multiplication of two commitments would be a commitment to the sum of the committed values. In detail,

$$com(x_1 + x_2, r_1 + r_2) = g^{x_1 + x_2} h^{r_1 + r_2} = g^{x_1} h^{r_1} \times g^{x_2} h^{r_2} = com(x_1, r_1) \times com(x_2, r_2)$$

### 4.2)    Schnorr signature scheme

Instead of running an inefficient zero knowledge proof, we will add a Schnorr signature to the Pedersen commitments for the central bank to verify that the transaction amount is preserved in a concealed transaction (with no over-issuance). Like the Pedersen commitment scheme, a Schnorr signature scheme works on multiplicative groups. A Schnorr signature scheme over a subgroup of $Z_p^*$ of order $q$ (i.e., $q$ divides $p - 1$) works as follows:

Choose $x$ in $Z_q^*$ as the private key, $y = g^x$ is public key.

For a message $m$ to be signed, the signer first chooses a random $k$ in $Z_q^*$.

He then computes:

1)    $t = g^k \bmod p$

2)    $e = H(t \| m)$, $m$ is the message to be signed, where $H(\ )$ is a **cryptographic hash function** from $\{0,1\}^*$ to $Z_q^*$.

3)    $s = k - xe \bmod q$

The signature is $(s, e)$.

If $H(g^s y^e \| m) = e$, then the signature is verified.

Proof:
$$g^s y^e = g^{k - xe} y^e$$

$$= \frac{g^k y^e}{g^{xe}}$$

$$= \frac{t y^e}{y^e}$$

$$= t$$

$$H(g^s y^e \| m) = H(t \| m) = e$$

**4.3)      Transaction value concealment based on Pedersen commitment**

We now show how to use a Pedersen commitment scheme to hide the transaction values of the inputs and outputs of a **UTXO** transaction. For a given **UTXO** transaction with inputs $X = \{x_i\}$ and outputs $Y = \{y_j\}$, where $x_i > 0$, $y_j > 0$ and $x_i, y_j << q$ for all $i$, $j$, a bank computes the Pedersen commitments $c_i$ and $c'_j$ for $x_i$ and $y_j$ respectively as follows.

The bank chooses random integers $r_i$, $r'_j$ in $Z_q$* for all $i, j$ and commits the value $x_i$, $y_j$ to $c_i = g^{x_i} h^{r_i}$ and $c'_j = g^{y_j} h^{r'_j}$. The concealed transaction is then formed by using $C = \{c_i\}$ as inputs and $C' = \{c'_j\}$ as outputs. Each of the outputs $c'_j$ is attached with a range proof $rp_j$ to prove that the respective output value $y_j$ is non-negative. While negative values do not make any sense in modular arithmetic, the range proof proves that the value is smaller than a certain value in $Z_q$, assuming a simple encoding from integers to $Z_q$, which represent any negative number of $x$ as $q - x$. Standard range proof protocols [29, 30] can be used and will not be discussed in detail in this work. Putting these together, a transaction is given by $tx = (C, C', RP)$ where $C = \{c_i\}$, $C' = \{c'_j\}$, $RP = \{rp_j\}$.

The bank then generates a Schnorr signature $\sigma_{Bank}(C \| C')$ using the random secrets $r_i, r'_j$ as follows:

The bank first calculates the sum of random $r_i$ used in a **UTXO** transaction minus the sum of random $r'_j$ used in the same **UTXO** transaction. Let the difference be $\alpha$, that is, $\alpha = \left( \sum_i r_i - \sum_j r'_j \right) \bmod q$. To generate the Schnorr signature, $\alpha$ is used as the private key for signing, and the public key for verification is then given by $\beta = h^{\alpha}$.

The bank first chooses random $k$ from $Z_q{}^*$

Let $t = h^k \bmod p$.

Let $e = H(t \| m)$, $m$ is the message to be signed.

Let $s = k - \alpha e \bmod q$.

The signature generated is $(s, e)$.

The bank then sends these commitment values as inputs and outputs of the **UTXO** transaction, together with all the $(x_i, r_i)$, $(y_j, r'_j)$, for the payer of the transaction to sign to authorize the payment. The payer can verify the correctness of the concealed **UTXO** transaction by opening all the commitment values $c_i, c_j'$ using $(x_i, r_i)$, $(y_j, r_j')$. If they are correct and agrees with what the payer intends to pay to the payee, the payer then adds his signature $\sigma_{payer}(C \| C')$ and sends it to the bank. The bank then sends the concealed **UTXO** transaction, together with the two signatures, to the central bank for recording at the central bank ledger. That is, the transaction sent to the central bank is:

$$\left(C, C', RP, \sigma_{Bank}(C \| C'), \sigma_{payer}(C \| C')\right), \text{ where } C = \{c_i\}, \ C' = \{c_j'\}, \ RP = \{rp_j\}.$$

To verify whether it is a valid transaction to update its ledger, the central bank first verifies the payer's signature to check if it is an authorized payment. Then, it checks all the range proofs, that is, all pairs of $(c_j', rp_j)$ are verified to ensure that the committed values are non-negative.

Then, the central bank verifies that no over-issuance of **rCBDC** occurs in the transaction. To verify that, the central bank computes $z$ in order to verify that the sum of the outputs is equal to the sum of the inputs of the **UTXO** transaction (without decrypting or opening the commitment values). The verification is run as follows.

The central bank first calculates

$$z = \frac{\prod\limits_{c_i \in C} c_i}{\prod\limits_{c_j' \in C'} c_j'} = \frac{g^{x_1} h^{r_1} \times g^{x_2} h^{r_2} \times ... \times g^{x_i} h^{r_i}}{g^{y_1} h^{r_1'} \times g^{y_2} h^{r_2'} \times ... \times g^{y_j} h^{r_j'}} = \frac{g^{\sum x_i} h^{\sum r_i}}{g^{\sum y_j} h^{\sum r_j'}} = g^{\sum x_i - \sum y_j} \times h^{\sum r_i - \sum r_j'}$$

If the sum of inputs is equal to the sum of outputs, $\sum x_i - \sum y_j = 0$,

then, $z = g^0 * h^v = h^v$, where $v = \sum r_i - \sum r_j'$.

Note that $z$ is equal to the public key $\beta = h^\alpha$ if $\sum x_i - \sum y_j = 0$. That is, $z$ can be used as the public key to verify the signature $\sigma_{Bank}(C \| C')$. If the signature verification is passed, the central bank can be assured that $\sum x_i - \sum y_j = 0$, that is, no over-issuance has occurred in the transaction.

When all the verifications pass, the central bank can confirm:

(1) no over-issuance occurs in the transaction;

(2) the transaction is authorized by the legitimate owner of **CBDC**.

The central bank, therefore, updates its ledger by purging all the concealed **UTXO**s corresponding to the concealed inputs of the current **UTXO** transaction and adding all the concealed outputs of the transaction to its ledger. This completes the update of the central bank ledger.

To confirm the transaction for the bank and payer, the central bank collects all the concealed outputs of the transaction and generates a Merkle tree [26] using these outputs as leaves. The central bank signs on the root of the Merkle tree and sends the signature back to the bank, which passes to a payee the respective concealed outputs, the corresponding random secrets needed to open the commitments, the internal node values of the Merkle tree off the paths of the leaves corresponding to the concealed outputs, and the central bank's signature of the confirmation on the root of the Merkle tree. With these values, the payee can verify whether the payment is valid, that is, has been processed and finalized, as well as, updated in the central bank ledger.

## 5)    Security Analysis of Transaction Concealment

In this section, we analyse how the proposed transaction concealment design can achieve the following security assurance:

1) No over-issuance of money: if the concealed **UTXO** transaction does not contain commitment values such that the sum of the outputs is equal to the sum of the inputs, the bank would not be able to create a signature $\sigma_{Bank}(C \| C')$ passing the verification by the central bank.

2) No leakage of any transaction value to the central bank: the concealed **UTXO** transaction and the associated data passed to the central bank and recorded on its ledger do not leak any information about the concealed transaction values to the central bank.

3) No unauthorized transaction unless the payer agrees: without the authorization of the owner of a given concealed **UTXO** (recorded on the central bank ledger), nobody can spend the **UTXO**. The bank will not be able to spend it either.

4) Fraudulent transaction detection: unless with the collusion of the central bank, nobody (including the bank) can deceive a payee into accepting a **UTXO** that has been spent or is not authorized by the respective **UTXO** owner.

## 5.1)    No over-issuance of money

<u>Claim 5.1</u>

If the discrete log problem is computationally difficult for the multiplicative group used to generate the Pedersen commitments, a malicious bank will not be able to generate a correct signature for a concealed **UTXO** transaction passing the central bank's verification unless the commitment values of the concealed transaction are commitment values for a set of inputs and outputs which satisfy that the sum of the outputs is equal to the sum of inputs.

Proof:

We prove this security property by contradiction.

Suppose a concealed **UTXO** transaction contains commitment values whose inputs and outputs do not satisfy that the sum of inputs is equal to the sum of outputs. That is, $c_i = g^{x_i} h^{r_i}$ and $c'_j = g^{y_j} h^{r'_j}$, but

$$U' = \sum x_i - \sum y_j \neq 0$$

To verify the transaction, the central bank computes $z'$ as follows:

$$z' = \frac{\prod\limits_{c_i \in C} c_i}{\prod\limits_{c'_j \in C'} c'_j} = \frac{g^{x_1} h^{r_1} \times g^{x_2} h^{r_2} \times ... \times g^{x_i} h^{r_i}}{g^{y_1} h^{r'_1} \times g^{y_2} h^{r'_2} \times ... \times g^{y_j} h^{r'_j}} = g^{\sum x_i - \sum y_j} \times h^{\sum r_i - \sum r'_j} = g^{U'} h^v$$

where $v = \sum r_i - \sum r'_j$.

Let $g = h^b$ for some unknown $b$.

Then $z' = h^{bU'+v}$. The central bank then uses $z'$ as the public key to verify the signature $\sigma_{Bank}(C \| C')$. To fool the central bank to accept $C, C'$, the adversarial bank has to generate a signature, which likely need the knowledge of the private key for $z'$. Here, we rely on the proof of [28], which assumes that the hash function $H$ used in the Schnorr signature scheme is a random oracle. By assuming $H$ as a

random oracle, it is possible to rewind the adversary to make it generate two different Schnorr signatures with non-negligible probability. With these two signatures, we can find the unknown private key for $z'$ with a certain, non-negligible probability. While [28] shows a proof for security against existential forgery and we consider selective forgery here, the proof of [28] also works here.

That is, the bank knows some $d = bU' + v$. But knowing $d$ will enable it to determine $b$ as follows:

$$b = \frac{d - v}{U'}$$

However, this enables the bank to find the discrete log of $g$ base $h$ (supposed to be difficult).

The Schnorr signature ensures that the sum of inputs of a **UTXO** transaction is equal to the sum of its outputs without the need to open the concealed commitments. However, if an adversary puts in negative transaction values to some of the outputs, the transaction could still pass this verification but over-issuance of **rCBDC** has occurred in the transaction. To withhold this attack, a range proof for each output of a **UTXO** is attached to prove that each of the values used to form the commitments is non-negative. Hence, if the range proof used is sound, it should not be possible to have negative values committed in any of the output commitment values.

### 5.2)    No information leakage to the central bank

Without loss of generality, we consider a simple concealed **UTXO** transaction with one input and two outputs and assume that the central bank is able to open the input commitment (i.e., the central bank knows the transaction value and the random secret used to create the concealed input). That the central bank knows the value and secret of the input commitment is a fair assumption since the central bank knows the transaction value of the first **UTXO** when a given amount of **CBDC** is first minted.

### <u>Claim 5.2</u>

If the Pedersen commitment scheme is perfectly hiding, the central bank is unable to obtain the transaction value of any one of the two concealed outputs.

Proof:

We prove this by contradiction.

Suppose the input transaction value is $x$, and we pick one of the outputs with transaction value $y_1$. Then the other output has a transaction value of $(x - y_1)$.

Let the commitment values for them be $c'_1 = g^{y_1} h^{r'_1}$ and $c'_2 = g^{(x-y_1)} h^{r'_2}$. Denote the commitment value for the input by $c = g^x h^r$. Assume that the central bank has a method **W** to find $y_1$ from $(c, c'_1, c'_2)$, $x$, $r$. It is possible to show that method **W** can be used to break the Pedersen commitment scheme as follows:

Given a commitment $c'_1 = g^{y_1} h^{r'_1}$, where $y_1$ and $r'_1$ are unknown, we can arbitrarily pick $x$ and $r$ to form $c = g^x h^r$. If we can compute $c'_2$, we have $(c, c'_1, c'_2)$, $x$, $r$ to feed to method **W** to find $y_1$.

What remain is how to compute $c'_2$ for feeding to **W**. We can compute $c'_2$ by computing the following:

$$w = \frac{c}{c'_1}$$

By substituting $c = g^x h^r$ and $c'_1 = g^{y_1} h^{r'_1}$ into $w$, we can see that $w$ is a commitment of $(x - y_1)$, i.e., equivalent to $c'_2$ if we take $r'_2 = \left(r - r'_1\right)$. The derivation is as follows:

$$w = \frac{c}{c'_1} = \frac{g^x h^r}{g^{y_1} h^{r'_1}} = g^{(x-y_1)} h^{(r-r'_1)} = g^{(x-y_1)} h^{r'_2}$$

This contradicts with the assumption that the Pedersen commitment scheme is hiding since we manage to find $y_1$ from its commitment $c'_1$. Through a contrapositive argument, if the Pedersen commitment scheme is secure with respect to message hiding, then method **W** cannot exist. That is, the central bank cannot find $y_1$ from the commitment values of a concealed transaction.

While the commitment values in the concealed inputs and outputs of the **UTXO** transaction do not leak information about the transaction values to the central bank, what about the range proofs and the attached signatures? For the range proof, based on the property of a zero knowledge proof, the transcript should not leak information about the transaction values. For the payer signature, since the message is on concealed commitment values and the public key is independent of the transaction values or commitment values, it should not contain information about the transaction values. For the Schnorr signature, the private key is derived from the random secrets which are also used for the commitment scheme. But for a Schnorr signature $(s, e)$, the first part should be indistinguishable from a random number. The second part e is a hash value; if the hash function $H$ is a random oracle, it should also be like a random number. In other words, they should leak no information about the transaction values.

### 5.3)     No unauthorized transaction

The central bank would verify the signature of the payer on a concealed **UTXO** transaction before updating its ledger. Note that the payer is the owner of the input **UTXO**s. If the signature scheme used by the payer is unforgeable, nobody other than the payer can create a valid signature to request the central bank to purge the input **UTXO**s. In other words, only the rightful owner of the input **UTXO**s of a transaction can make the central bank process his transaction, and no unauthorized transaction will be processed by the central bank to update its ledger, which is the ultimate reference for every party.

### 5.4)     Fraudulent transaction detection

If an attacker impersonates the owner of a **UTXO** to create a fraudulent transaction to deceive a payee, he will not be able to create the required signature to request the central bank to process his transaction. That is, the central bank will not create its confirmation signature on the transaction. Without the central bank's signature, the payee will know that the transaction is unconfirmed or not processed.

In case an owner of a **UTXO** double spends it, that is, the same **UTXO** is used as input for more than one transaction to spend it multiple times, the central bank would detect it and only process the first received transaction from the owner or payer. Recall that the central bank purges input **UTXO**s of a transaction from its ledger when the transaction is processed. When a second transaction spending the same **UTXO** is received by the central bank, the said **UTXO** should have been erased from the central bank's ledger, and the central bank will not sign to confirm the transaction. The second transaction will be invalidated.

### 6)     Transaction Aggregation through Coinjoin

In this section, we show how multiple concealed **UTXO** transactions from different payers can be merged by the bank into a larger transaction via Coinjoin while preserving the privacy properties achieved in Section 4. Shown in Figure 4 is transaction aggregation via Coinjoin. The key is to dissolve the existing transaction boundaries. Since inputs and outputs of a **UTXO** transaction can be mixed and shuffled freely, they would not pose any problem in aggregation. However, what message is being signed by a signature is essential for signature verification, and the transaction boundaries need to be imposed for correct signature verification. However, this will disclose the original transaction boundaries, and anonymization through transaction aggregation is not achievable. To achieve aggregation, it is necessary to ask all the payers to sign on the aggregated transaction instead.

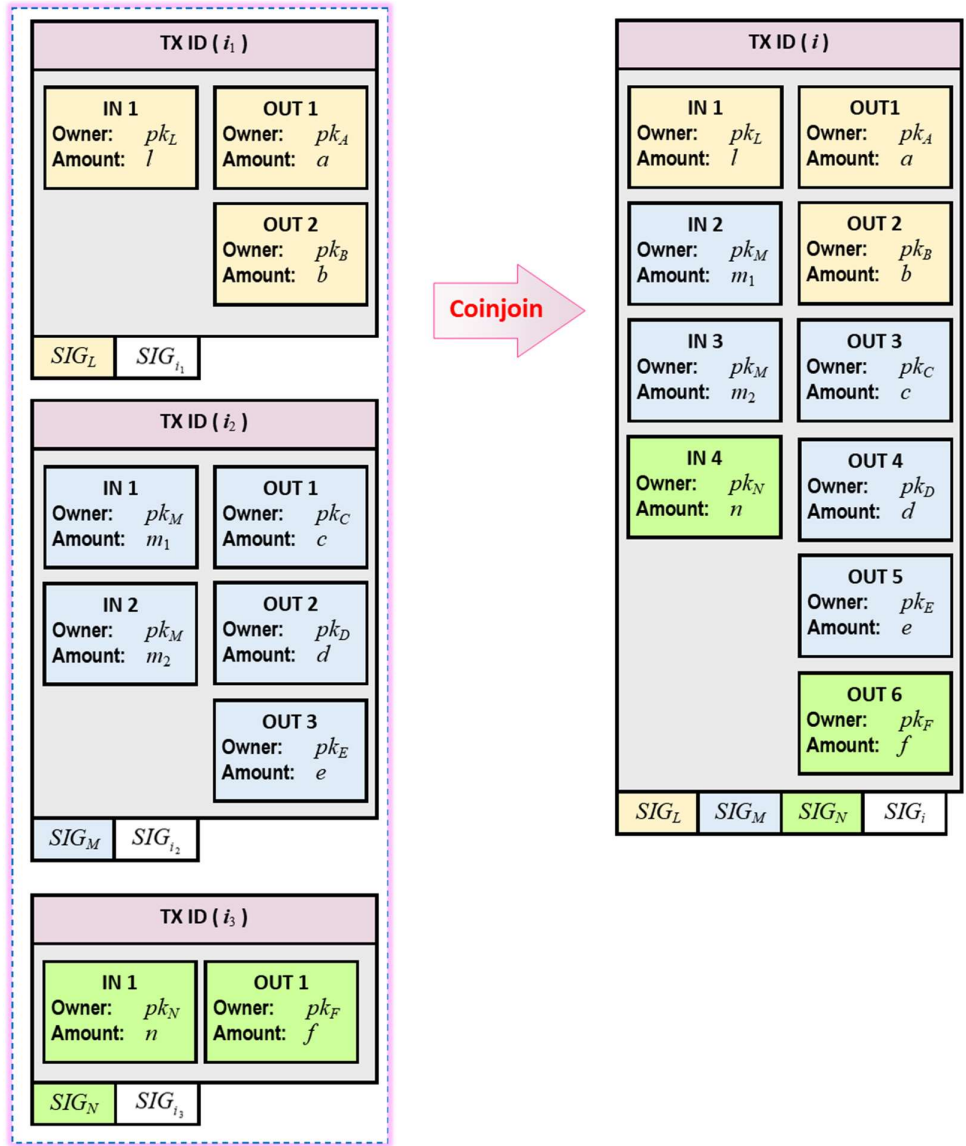

Figure 4. Aggregation of concealed **UTXO** transactions via Coinjoin

For the Schnorr signatures used to prove that no over-issuance occurs in the transactions before merging, they are generated by the bank, which is ready to create a signature for the aggregated transaction instead. For payers, the bank would need to ask them to sign on the aggregated transaction, and the signers may doubt if the transaction would deceive them into signing an unauthorized transaction to spend their money. To solve this issue, the bank can extract a payer's transaction from the aggregated transaction and send it along with the aggregated transaction to the payer. Note that a copy of the random secrets used to generate the Pedersen commitments for the payer's transaction will be sent by the bank and kept by the payer and payee (according to Section 4). By checking all the inputs in the aggregated transaction under his public key, the payer can be assured that the aggregated transaction does not spend extra **UTXO**s from him. The payer can then check whether all his intended outputs are included in the **UTXO** transaction. If they are, that means his **rCBDC** will only be paid to his intended payees only. Hence, the payer can be convinced that the aggregated transaction is not a fraudulent transaction and would confidently sign on it.

### 6.1)    Anonymity of an aggregated transaction through Coinjoin

In an aggregated transaction, there are multiple payers and multiple payees, and it is not possible to decide who pays whom since the transaction inputs and outputs are hidden and concealed in the Pedersen commitment values, which look like truly random numbers. That is, the payer-payee relationships in the group are obfuscated. For a given payee, one cannot tell for sure which of the payers has paid him. Similarly, for a given payer, one cannot ascertain, from the transaction, to whom he has paid. This, therefore, achieves some form of counterparty anonymity. To properly characterize this anonymity property, we leverage a well-developed notion in database research, namely, the $k$-anonymity, and define counterparty anonymity as follows:

**Definition 6.1 ($k$-counterparty-anonymity)**

A transaction is said to satisfy $k$-counterparty-anonymity (or simply $k$-anonymity) if each payee (or payer) in the transaction can be indistinctly matched to at least $k$ possible payers (or payees) as his counterparty.

---

***Example 6.2*** In the example shown in Figure 4, looking at the transaction alone, each payee can possibly be thought of as transacting with one of the three payers ($L$, $M$ or $N$), and no one can be certain which is the actual payer to a given payee. Similarly, each payer can be thought of as transacting with one of the six payees ($A$, $B$, $C$, $D$, $E$ or $F$). Taking the smaller of the two numbers, the transaction satisfies 3-anonymity.

---

Now, the problem of aggregating transactions from multiple signers is that the first payer has to wait for the arrivals of other transactions from other payers before his transaction can be confirmed. In general, to achieve better anonymity (i.e., a greater $k$), a longer waiting time for transaction confirmation is necessarily incurred.

**6.2)    Trade-off between anonymity and transaction confirmation time**

In order to analyse the waiting time needed to achieve $k$-anonymity for aggregated transactions, we model the arrivals of transactions as a Poisson process with a rate parameter $\lambda$ (shown in Figure 5). For the sake of simplicity, let us assume each transaction consists of at least one payee which is not the same as the payer.

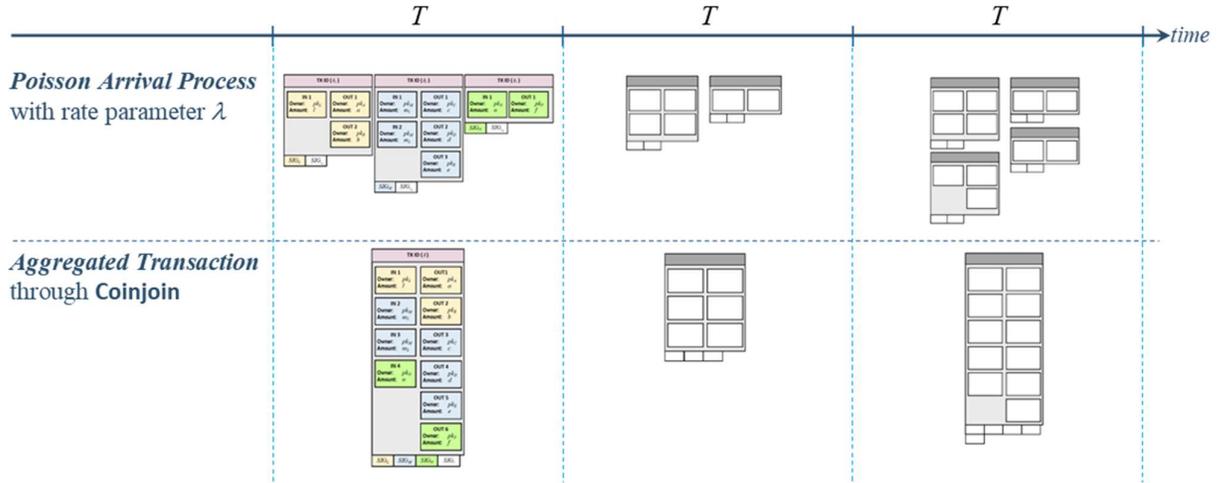

Figure 5.  Aggregating transactions with arrivals following a Poisson distribution

For a Poisson process for transaction arrivals, the probability that $k$ transaction arrivals occur in a finite interval $T$ is given by:

$$p(k) = \frac{(\lambda T)^k e^{-\lambda T}}{k!} \text{, where } k = 0, 1, 2, \ldots$$

where $\lambda$ is the rate parameter.

According to [27], it can be easily shown that the mean and variance is respectively given by:

$$E(k) = \lambda T \text{ and } \mathrm{var}(k) = \lambda T$$

_Level of anonymity achieve for fixed $T$_

Given that transaction arrivals follow a Poisson process with a rate parameter $\lambda$, if a certain time interval $T$ is waited to aggregate transactions from different users into a single transaction, the probability to achieve $k$-anonymity is given in Figure 6.  The plot is actually an inverse cumulative distribution function (CDF) of the Poisson process (i.e., $1 - \Pr\left[K \leq k\right]$), for different values of $\lambda T$.

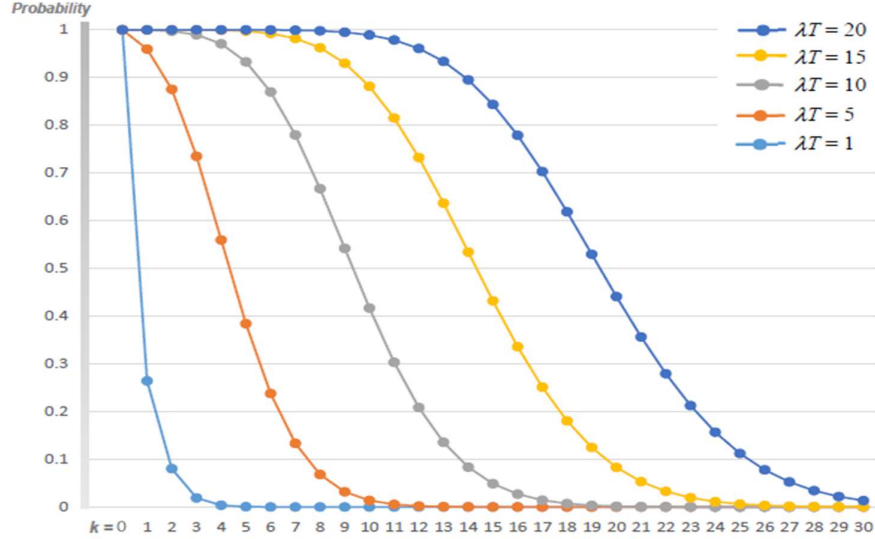

Figure 6.  Probability to achieve *k*-anonymity for Poisson transaction arrivals

As can be seen, for a given $\lambda$, as $T$ increases, it is more likely to achieve greater anonymity since a greater *k* for *k*-anonymity can be achieved.  As a rough estimate, we can determine the achieved level of anonymity from the rate parameter $\lambda$ and waiting time $T$ as follows:

$$k(\text{achieved}) = \max(E(k) - 2\sigma_k, 0) = \max(\lambda T - 2\sqrt{\lambda T}, 0)$$

The preceding formula serves as a good estimate to determine the achieved *k* for a given transaction arrival process with rate parameter $\lambda$ if transaction aggregation is performed periodically with a constant period $T$.  For the given $\lambda$, $T$, transaction aggregation through Coinjoin can achieve *k*-anonymity with reasonably high probability (> 99%) according to Figure 6.

### _Waiting time required to achieve k-anonymity_

In practice, it may be necessary to determine the waiting time needed to achieve *k*-anonymity for a fixed *k* with reasonably high probability.  In the following, we will derive the waiting time needed to achieve *k*-anonymity given the transaction arrival rate $\lambda$.

It is a well-known result [27] that the inter-arrival time (i.e., the time between the arrivals of the 2 consecutive transactions) of a Poisson process follows an exponential distribution with the probability distribution function given by:

$$f_X(x) = \lambda e^{-\lambda x}, \quad x \in [0, \infty)$$

When a transaction $tx_0$ is received, it has to wait for the arrival of $(k-1)$ other transactions $tx_i$ for $i = 1, \ldots, k-1$ in order to achieve $k$-anonymity.

Let $k' = k - 1$.

Let the arrival time for these $k'$ transactions be the random variable $X_i$, $i = 1, 2, \ldots, k'$.

As a property of the Poisson process, these $X_i$'s should follow an identical, independent distribution $(i, i, d)$. The waiting time to achieve $k$-anonymity is a random variable $Z = \sum_{i=1}^{k'} X_i$, where each $X_i$ follows the exponential distribution,

$$f_{X_i}(x_i) = \lambda e^{-\lambda x_i}, \quad \text{where } x_i \in [0, \infty)$$

**Theorem 6.3:** The waiting time to achieve $k$-anonymity $Z$, given that the transaction arrival process is a Poisson process with rate parameter $\lambda$, is a gamma distribution $\Gamma(\lambda, k')$ with parameter $\lambda$, $k'$, i.e.,

$$f_Z(z) = \frac{\lambda e^{-\lambda z} (\lambda z)^{k'-1}}{\Gamma(k')}, \text{ where } \Gamma(k') = (k'-1)! \text{ for } k' > 0$$

Proof:

Let $X \sim \Gamma(\lambda, s)$ and $Y \sim \Gamma(\lambda, t)$ be two independent random variables following the gamma distribution. Note that the exponential distribution is a special case of gamma distribution, that is, $\Gamma(\lambda, 1)$. We will show that the sum of two random variables, each independently following a certain gamma distribution, is also a gamma distribution. $X$ is the total time of seeing $s$ events, while $Y$ is the amount of subsequent time till seeing $t$ more events.

$$f_X(x) = \frac{\lambda e^{-\lambda x} (\lambda x)^{s-1}}{\Gamma(s)} \text{ and } f_Y(y) = \frac{\lambda e^{-\lambda y} (\lambda y)^{t-1}}{\Gamma(t)}$$

The probability density function of the sum of random variables is given by:

$$f_{X+Y}(a) = \int_{-\infty}^{\infty} f_X(a-y) f_Y(y) dy$$

Substituting $f_X(x)$ and $f_Y(y)$, we have:

$$f_{X+Y}(a) = \int_0^a e^{-\lambda(a-y)} (a-y)^{s-1} e^{-\lambda y} y^{t-1} dy = e^{-\lambda a} \int_0^a (a-y)^{s-1} y^{t-1} dy$$

As $\int_0^a (a-y)^{s-1} y^{t-1} dy$ is a power of $a$, let $x = \frac{y}{a}$, the equation becomes:

$$\int_0^1 (a - ax)^{s-1} (ax)^{t-1} (a \, dx) = a^{s+t-1} \int_0^1 (1-x)^{s-1} x^{t-1} dx$$

So, $f_{X+Y}(a) = e^{-\lambda a} \cdot a^{s+t-1} \int_0^1 (1-x)^{s-1} x^{t-1} dx = \dfrac{\lambda e^{-\lambda a} (\lambda a)^{s+t-1}}{\Gamma(s+t)}$

We can then conclude that $(X + Y) \sim \Gamma(\lambda, s + t)$. That is $(X + Y)$ also follows a gamma distribution.

Note that $\Gamma(\lambda, 1)$ is an exponential distribution. The sum of two random variables $X_1, X_2$ each following an exponential distribution is a random variable following $(X_1 + X_2) \sim \Gamma(\lambda, 2)$. By standard mathematical induction, we can easily prove that $Z = \sum_{i=1}^{k'} X_i \sim \Gamma(\lambda, k')$.

Figure 7 shows the CDF of $f_Z(z)$ for different $\lambda$ and $k$. The waiting time can be looked up from the plot. For example, to achieve a 99.9% probability for $\lambda = 1$, $k = 20$, the time needed is 31.

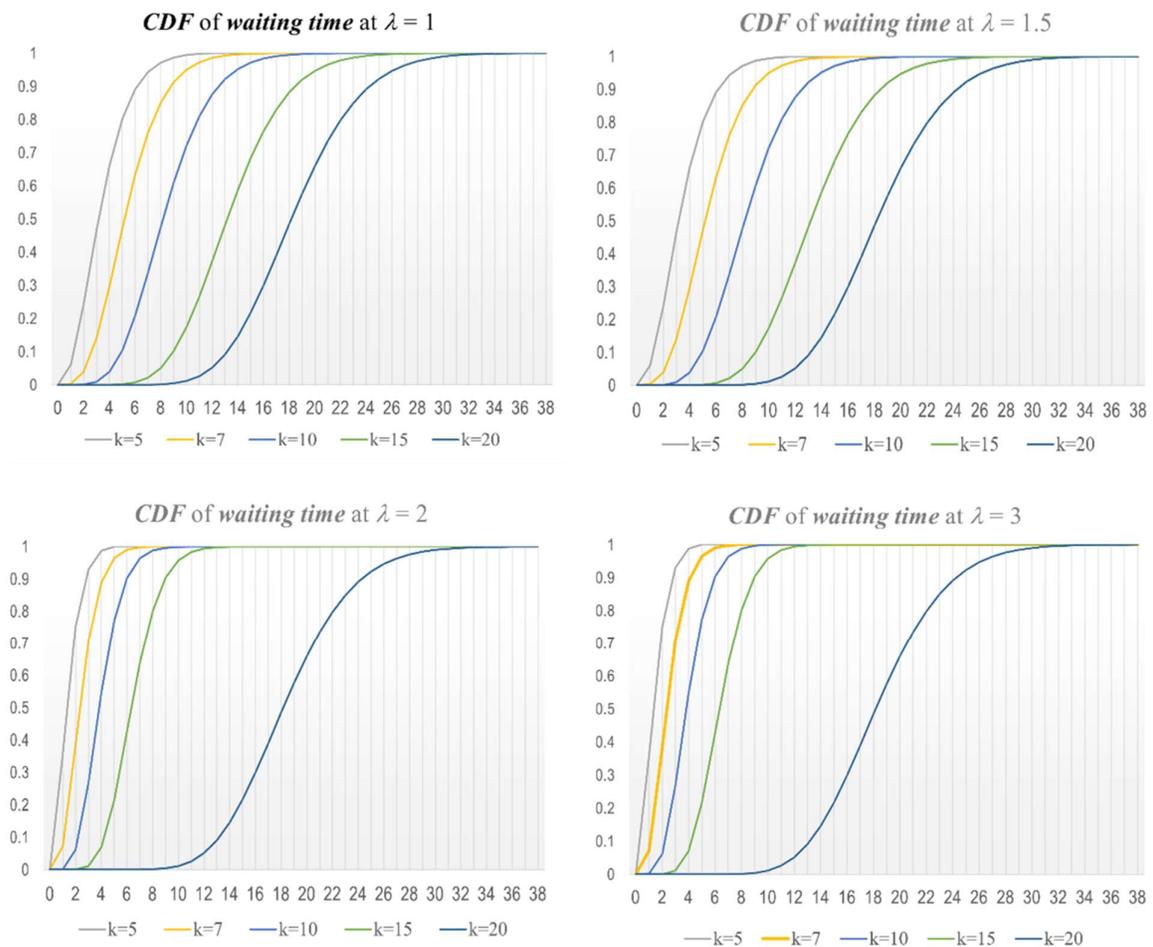

Figure 7.  CDF for the waiting time to achieve $k$-anonymity for different $\lambda$'s

## 7)    Performance Analysis and Simulation Results

### 7.1)    Overheads of concealing UTXO transactions

Assume a multiplicative subgroup of $Z_p^*$ is used for both the Pedersen commitment scheme and Schnorr signature scheme and $p$ is $k$ bits long. Suppose $q$ **UTXO** transactions are aggregated to give an aggregated transaction with $m$ inputs and $n$ outputs, excluding the overheads of the range proof.

*Computation overhead by bank:*

To generate a Pedersen commitment, 2 exponentiations and 1 multiplication is carried out,

$$\text{Computational overhead} = O(2k^3) + O(k^2) = O(2k^3)$$

To generate the commitment for an aggregated transaction with $m$ inputs and $n$ outputs,

$$\text{Computational overhead} = (m+n) \times O(2k^3) = O\left(2(m+n)k^3\right)$$

To generate a Schnorr signature, 1 exponentiation, 1 multiplication, 1 addition and 1 hash is carried out,

$$\text{Computational overhead} = O(k^3) + O(k^2) + O(k) + O(1) = O(k^3)$$

To generate the signing key for an aggregated transaction with $m$ inputs and $n$ outputs,

$$\text{Computational overhead} = (m+n-1)O(k)$$

Total computational overhead for generating the Schnorr signature scheme $= O(k^3 + (m+n-1)k)$

$\therefore$ Total computation overhead by bank $\begin{aligned}[t] &= O\left(2(m+n)k^3\right) + (k^3 + (m+n-1)k) \\ &= O\left(k^3(2m+2n+1) + (m+n-1)k\right) \\ &= O\left(k^3(2m+2n+1)\right), \text{ assuming } k >> m,n \end{aligned}$

*Computation overhead by payer:*

Total computation overhead by payer for signature $= O(k^3)$

*Computational overhead by central bank:*

To verify the Pedersen Commitments, $(m+n-1)$ multiplications are carried out,

$$\text{Computational overhead} = (m+n-1)O(k^2) = O\left((m+n-1)k^2\right)$$

To verify the Schnorr signature, 2 exponentiations, 1 multiplication and 1 hash is carried out,

$$\text{Computational overhead} = O(2k^3 + k^2 + 1)$$

$\therefore$ Total computational overhead by central bank $\begin{aligned}[t] &= O\left((m+n-1)k^2 + 2k^3 + k^2 + 1\right) \\ &= O\left(2k^3 + (m+n)k^2 + 1\right) \\ &= O(2k^3), \text{ assuming } k >> m,n \end{aligned}$

**7.2)       Design trade-off of transaction aggregation**

By merging different **UTXO**s together, we can increase the level of anonymity (i.e., $k$), but at the same time, this increases the processing time for confirming transactions. Simulation is run to determine the waiting time for achieving $k$-anonymity for different values of the rate parameter $\lambda$ of the Poisson transaction arrival process. We use Google Colab to carry out the simulation, and the code for simulation is in Appendix.

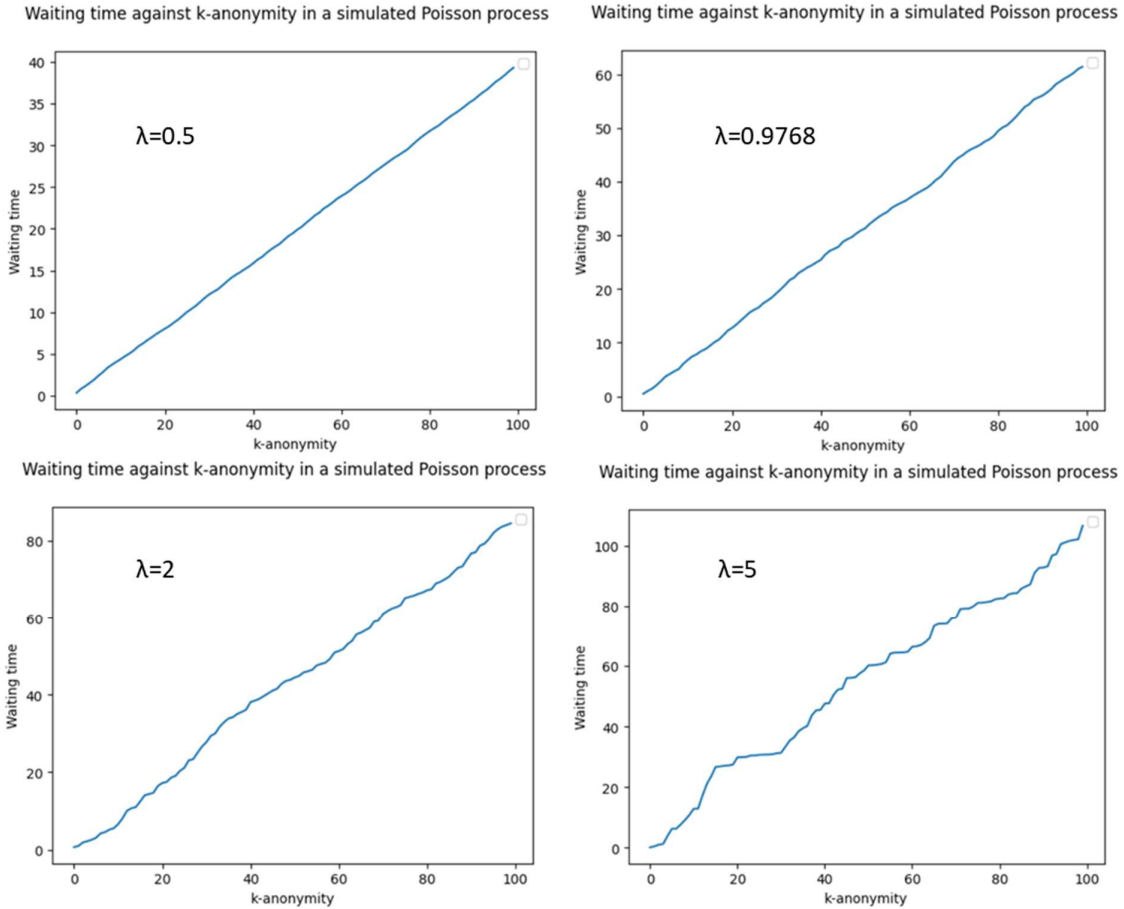

The graphs above show the simulation results of waiting time against $k$-anonymity using Poisson processes with different $\lambda$'s. As can be seen, regardless of the value of $\lambda$, the waiting time has roughly a linear relationship with the level of anonymity. We can therefore conclude that the trade-off between the level of anonymity and the time required for transaction confirmation is linearly related. That is, the greater the level of anonymity required, the longer the waiting time required. While the waiting time proportionately increases if a higher anonymity is required, the relationship is not strictly linear at large $\lambda$'s, possibly attributed to a larger variance for a larger $\lambda$.

## 8)    Conclusion

While **rCBDC** is widely seen as a key upgrade of the monetary system for the 21$^{st}$ century, privacy concern is likely the key impediment to the development and roll-out of **rCBDC**. While many central banks tried to mitigate the privacy issue through not keeping retail transaction records on the central bank ledger, this makes it difficult for the central bank to detect and prevent over-issuance and introduces other problems. This work aims to leverage technology to address the privacy problem of **rCBDC**. While pseudonym schemes and evolving public keys have been proposed to enhance user anonymity, gaps still exist in user privacy protection. This work addresses these gaps through the use of cryptographic protocols and a privacy enhancing technology called Coinjoin. The Pedersen commitment scheme combined with a Schnorr signature is used to hide transaction values of a **UTXO** transaction from the central bank while allowing the central bank to verify that the amount of inputs and outputs of a transaction is preserved (i.e., no over-issuance). The proposed protocol thus strikes the right balance between protecting user privacy and enabling central banks to fulfill their mandates. Coinjoin is used to merge multiple transactions from different payers into a larger transaction to obfuscate the payer-payee relationships of a transaction. In order to analyse the level of anonymity achieved through Coinjoin, this work proposes using $k$-anonymity, a well-known concept in database research, to evaluate transaction aggregation. The trade-off between the level of anonymity and the transaction confirmation time is also illustrated.

As future work, a number of areas could be further improved. First, the range proof used in the protocol still rely on inefficient zero knowledge proof. Further exploration to enhance the range proof should be pursued. Besides, the current simulation does not take into consideration that multiple transactions in the same confirmation window could possibly come from the same payer or be sent to the same payee, thus lowering the level of anonymity achieved. Future work should include a probability model to account for such transactions with the same payer or payee.

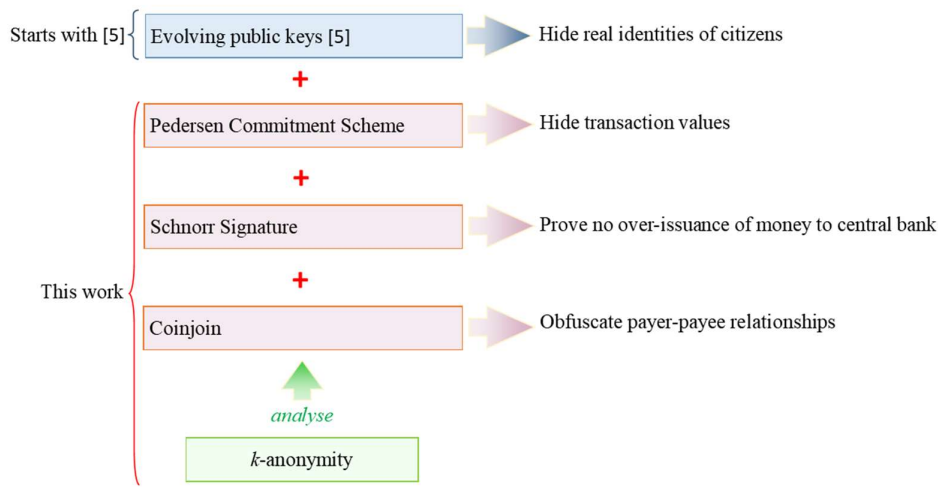

## Acknowledgement

The author is especially indebted to Mr. Wong Wai Hung for closely guiding this work.
The author also thanks Dr. Aldar Chan for suggesting the topic and providing technical suggestions to the work.

**Appendix: Code used in simulation**

```python
import random
import math
import matplotlib.pyplot as plt

_lambda = 0.9768
_total_anonymity = 100
_k_anonymity = []
_inter_waiting_times = []
_waiting_times = []
_waiting_time = 0
print('k_anonymity,inter_waiting_time,waiting_time')

for i in range(_total_anonymity):
  _k_anonymity.append(i)
  n = random.random()

  _inter_waiting_time = _lambda*math.exp(-1*_lambda*n)
  _inter_waiting_times.append(_inter_waiting_time)
  _waiting_time = _waiting_time + _inter_waiting_time
  _waiting_times.append(_waiting_time)
  print(str(i) +',' + str(_inter_waiting_time) + ',' + str(_waiting_time))

fig = plt.figure()
fig.suptitle('Waiting time against k-anonymity in a simulated Poisson process')
plot, = plt.plot(_k_anonymity, _waiting_times)
plt.legend(handles=[plot])
plt.xlabel('k-anonymity')
plt.ylabel('Waiting time')
plt.show()
```